\documentclass[journal, twocolumn, 18pt]{IEEEtran}

\usepackage{graphicx}
\usepackage{amsmath}
\usepackage{amssymb}
\usepackage{stfloats}

\normalsize

\ifCLASSINFOpdf

\else

\fi

\begin{document}

\title{Blockchain-Based Approach for Securing Spectrum Trading in Multibeam Satellite Systems}

\author{Feng Li,~\IEEEmembership{Member,~IEEE,}
        Kwok-Yan Lam,~\IEEEmembership{Senior Member,~IEEE,}
        Min Jia,~\IEEEmembership{Senior Member,~IEEE,}
        Jun Zhao,~\IEEEmembership{Senior Member,~IEEE,}
        Xiuhua Li,~\IEEEmembership{Member,~IEEE,}
        and Li Wang

\thanks{This research is supported by the National Research Foundation, Prime Minister's Office, Singapore under its Strategic Capability Research Centres Funding Initiative. Also, this work was supported by the Natural Science Foundation of Zhejiang Province under Grant LY19F010009 and LY19F010008. }

\thanks{F. Li is with School of Information and Electronic Engineering, Zhejiang Gongshang University, Hangzhou, 310018, China. F. Li is also at School of Computer Science and Engineering, Nanyang Technological University, 639798, Singapore. (fengli2002@yeah.net)}

\thanks{K. Lam and J. Zhao are with School of Computer Science and Engineering, Nanyang Technological University, 639798, Singapore. (kwokyan.lam@ntu.edu.sg, junzhao@ntu.edu.sg)}

\thanks{M. Jia is with School of Electronics and Information Engineering, Harbin Institute of Technology, Harbin, 150080, China. (jiamin@hit.edu.cn) }%

\thanks{X. Li is with School of Big Data \& Software Engineering, Chongqing University, Chongqing 401331, China. (lixiuhua1988@gmail.com)}

\thanks{L. Wang is with College of Marine Electrical Engineering, Dalian Maritime University, Dalian, 116026, China. (liwang2002@dlmu.edu.cn)}

}

\markboth{}%
{Submitted paper}

\maketitle

\begin{abstract}
This paper presents a blockchain-based approach for securing spectrum sharing in multi-beam satellite systems. Satellite spectrum is a scarce resource that requires highly efficient management schemes for optimized sharing by network users. However, spectrum sharing is vulnerable to attacks by malicious protocol participants. In order to ensure efficient spectrum management in the face of dishonest satellite users or cyber attackers, it is important for spectrum sharing mechanism to  provide transparency and traceability of the trading process so as to enable the system to detect, and hence eliminate, unauthorized access by malicious users.  We address these requirements by proposing the use of blockchain which, apart from its ability to provide transparency and traceability, ensures an immutable means for keeping track of user trading reputation. Besides, in order to address the practical constraints of heterogeneous user nodes, we also propose the use of edge computing to support users with limited computing power. In this paper, we propose a blockchain-based spectrum trading framework and, based on which, a multibeam satellite spectrum sharing algorithm for interference pricing and heterogeneous spectrum demands is devised to improve the efficiency of satellite spectrum. By leveraging on the system characteristics of blockchain, a dynamic spectrum sharing mechanism with traceability, openness and transparency for whole trading process is presented. Numerical results are also provided to evaluate the system benefits and spectrum pricing of the proposed mechanism.
\end{abstract}



\IEEEpeerreviewmaketitle

\section{Introduction}

\IEEEPARstart{T}{he} openness and wireless characteristics of satellite communications make it more vulnerable to security  interference by malicious users. For example, when terrestrial users share the spectrum with the satellite systems and access the satellite channel dynamically, the presence of malicious users will jeopardize any spectrum management effort of the satellite system. Satellite communications have the characteristics of long communication distance, high communication quality and wide coverage, which has become one of the important means of modern communication [1]-[3]. However, many of these characteristics also bring security challenges to communication system.

Due to the importance of security in satellite communications, a lot research works have been conducted to address malicious interference and reconnaissance for satellite communications, including space technology based on adaptive antenna, spectrum expansion technology, on-board processing and amplitude limiting technology [4]-[6].
In recent years, with the growing demand of dynamic spectrum sharing and access in satellite systems, privacy-preservation and security issues receive much extensive attention of the research community [7]-[9]. In dynamic spectrum utilization of satellite communications, security concerns need to be tackled in spectrum sensing, spectrum allocation and spectrum switching. Similar to the scenario of dynamic spectrum access and sharing in terrestrial communications, malicious user behavior in the spectrum sensing process may lead to prolonged allocation of communication channel. Besides, malicious collusion to manipulate spectrum pricing distribution may also undermine the interests of the satellite communication systems.

While improving the efficiency of spectrum utilization under the assumption that all users honestly execute the spectrum trading protocol, a practical spectrum management system must also address abnormal situations when there are dishonest users. In this case, dishonest users may send malicious messages to disturb the spectrum trading process in order to manipulate spectrum prices, hence resulting in inefficient allocation of the previous satellite spectrum resource. Existing research work on satellite spectrum sharing security mainly focuses on the field of spectrum sensing [10][11]. {The issues of malicious collusion in the process of spectrum trading, which damages the interests of satellite systems, and malicious long-term monopoly in dynamic spectrum access are critical to the efficient operation of satellite systems and deserve in-depth investigation. Nevertheless, to date, very little research work on security and privacy preservation of spectrum pricing and transaction in satellite systems has been reported. Recently, the technology of blockchain has attracted growing attention from researchers in wireless communications to secure network security [12]-[16]. In [12], the authors proposed a blockchain-based verification protocol to enable and secure dynamic spectrum sharing in mobile cognitive radio networks. In [13], the authors explored the application of blockchain technology on dynamic spectrum sharing. [14] aims to solve the security issue by introducing a novel privacy-preserving dynamic spectrum trading mechanism on the basis of blockchain-based method. In [15], a blockchain-based spectrum sharing mechanism which can efficiently share idle spectrum in dense networks was proposed. In [16], a blockchain-based platform was raised to improve the spectrum sharing by introducing the spectral token which validates and tracks the use of a licensed spectrum to facilitate the dynamic use for secondary users. Yet, none of the published research work considered blockchain-based technology as a means for solving the issue of spectrum trading security in satellite systems.}


{This paper proposes a blockchain-based approach to secure spectrum sharing in multibeam satellite systems. We design a blockchain-based framework, based on which, an efficient spectrum management scheme which is capable of withstanding malicious participants is developed. By  leveraging on the system characteristics of blockchain, the proposed scheme provides transparency and traceability of the trading process. }

In this article, to secure the satellite spectrum sharing, the concepts of edge node computing and user trading reputation are introduced to build the blockchain wherein terrestrial base station or fusion center acts as the edge node or blockchain miner who needs to complete mining task, creation of new block and exchange of spectrum coin. Edge node's historical information and transaction record will be gathered to form the its trading reputation so as to decline the low computation-power or malicious user's access. On the basis of the secure sharing mechanism, a satellite spectrum trading method is raised by taking into account inter-cell interferences and satellite users' stochastic demands on diverse satellite spectrum to optimize satellite systems' benefits. Numerical results are provided to evaluate the spectrum sharing method's performances in terms of satellite profits and interference pricing. The reminder of this paper is as follows: The system model and framework of blockchain-based spectrum trading are given in Section II. The joint interference pricing and power control algorithm is presented in Section III. Then, we show the simulation part in Section IV. The conclusion of this article is given in Section V.

\section{System Model and Blockchain-Based Spectrum Trading Framework}

\subsection{System Model for Multibeam Satellite Communications}

\begin{figure}
  \begin{center}
	\includegraphics[height=60mm,width=94mm]{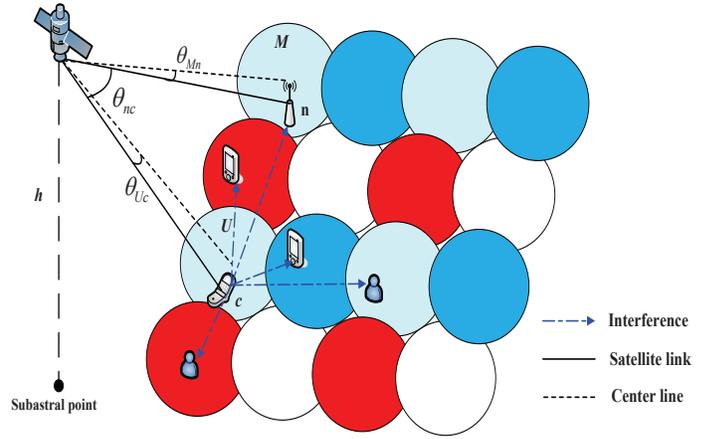}
	\caption{Spectrum reuse in multibeam satellite systems}
	\label{fig:dummy-figure}
  \end{center}
\end{figure}

In this paper, we consider the spectrum trading is carried out in multibeam satellite systems in which oblique projection and spectrum reuse mode are adopted as shown in Fig. 1. In this case, due to the inter-cell interference caused by spectrum reuse, we focus the interference pricing and power control when designing the spectrum sharing scheme.

In specific, for the effects of oblique projection, the angle $\theta_{Mn}$ describing the deviation angle between user $n$ and center point of cell $M$ can be expressed as
\begin{equation}\begin{split}
\theta_{Mn}=&\arccos\big(\{(d_o^s)^2+(d_{Mn}^s)^2-2R^2[1-\cos(d_{Mn}^o/R)]\}  \\
&\times(2d_o^s d_{Mn}^s)^{-1}\big),
\end{split}\end{equation}
where $d_o^s$ denotes the distance between cell center and the satellite. $d_{Mn}^s$ denotes the distance between user $(M, n)$ and the subastral point, $d_{Mn}^o$ denotes the distance between user $(M,n)$ and cell center, and $R$ means the earth radius. {Besides, the spectrum reuse and inter-cell interference in multibeam satellite systems are considered. As shown in Fig. 1, we assume user $n$ at cell $M$ will suffer interference from user $c$ at cell $U$ where the cells in same color share same spectrum band. Then, for the uplink channel, the receiving power at the satellite from user $n$ at cell $M$ can be expressed as}
\begin{equation}
P_r=\frac{p_{n}g_{n}(\alpha_{n})G_M(\theta_{n}^M)}{(t\pi d_{n}/\lambda)^2f_{n}(\alpha_{n})},
\end{equation}
{where $p_{n}$ is the transmit power of satellite user $n$, $g_{n}(\alpha_{n})$ is the antenna gain of satellite user $n$ at direction $\alpha_{n}$. $\theta_{n}^M$ denotes the derivation angle of user $n$ to the central line of cell $M$. $G_M(\theta_{n}^M)$ denotes the antenna gain of satellite cell $M$ in direction $\theta_n^M$. $d_n$ denotes the straight-line distance between $n$ and the satellite system. $\lambda$ is the wavelength, and $f_n(\alpha_n)$ is the channel fading for user $n$ in direction $\alpha_n$. Besides, inter-cell interference during the spectrum reuse can be expressed as }
\begin{equation}
I=\sum_{U=1}^k{\frac{p_{c}g_{c}(\alpha_{c})G_U(\theta_{c}^U)}{(4\pi d_{c}/\lambda)^2f_{c}(\alpha_{c})}\mu_{c}\rho_{U}^M}
\end{equation}
{where $\mu_{c}$ denotes the active factor of user $c$ at cell $U$ which is related to the user's service type. $\rho_U^M$ is the polarization isolation factor between cell $M$ and $U$.}

\begin{figure*}
  \begin{center}
	\includegraphics[height=90mm,width=180mm]{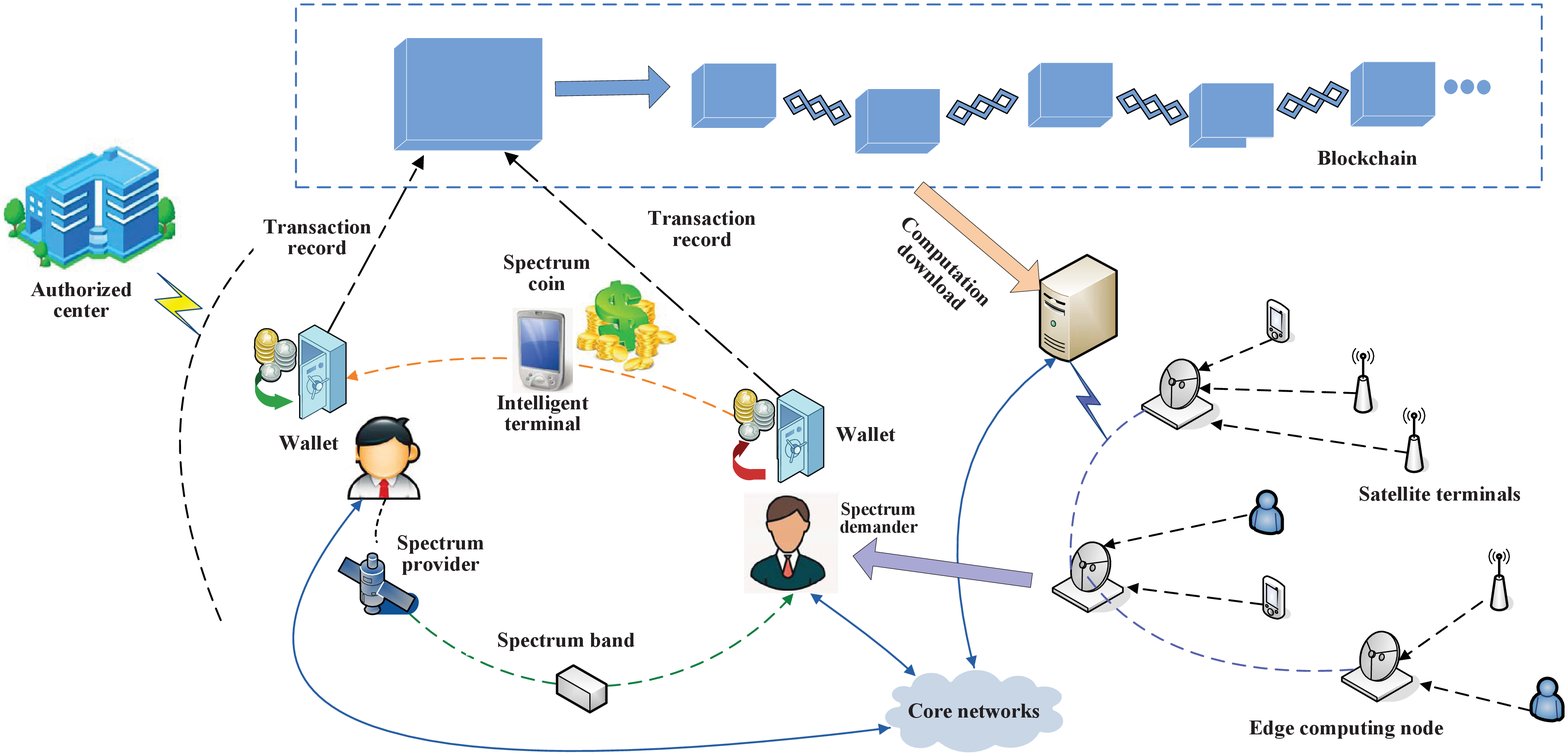}
	\caption{Blockchain-based satellite spectrum trading}
	\label{fig:dummy-figure}
  \end{center}
\end{figure*}

Then, for satellite user $(M,n)$, the transmission capacity with unit bandwidth can be expressed as
\begin{equation}\begin{split}
&C_{Mn}= \log_2(1+ \\
&\frac{\rho_{Mn}g_{Mn}(\varepsilon_{Mn})G_M(\varphi_{Mn})}{\!d_{Uc}^2 f_{Mn}(\varepsilon_{Mn})\!\sum\limits_{U=1}^l{\frac{\rho_{Uc}g_{Uc}(\varepsilon_{Uc})G_U(\varphi_{Uc})\mu_{Uc}\rho_{U}^M}{(4\pi d_{Uc}/M)^2f_{Uc}(\varepsilon_{Uc})}}+\!N_0(\varepsilon_{Mn})}),
\end{split}\end{equation}
{where $\rho_{Mn}$ is the transmit power of user $(M,n)$, $N_0(\varepsilon_{Mn})$ denotes the noise, $g_{Mn}$ and $G_M$ denote the antenna gain and $d_{Uc}$ denotes the straight-line distance from user $Uc$ to the satellite system. User $Uc$ locating in adjacent cell shares the same band with $Mn$}.

\subsection{Blockchain-Based Satellite Trading Framework}

When terrestrial users are allowed to share the spectrum with satellite systems and access satellite channels dynamically, or transmit the signal with the assistance of satellite systems, if the users are malicious, the security risk of the satellite systems will be obviously exposed. In addition, the malicious buyer, malicious third party and malicious collusion in the process of spectrum auction will also trigger confusion to satellite spectrum sharing. How to deal with the malicious user's access to satellite spectrum is a key issue to establish the secure spectrum sharing mechanism in multibeam satellite systems. This paper plans to make full use of the natural security attributes of blockchain technology to build a secure architecture for spectrum sharing in multibeam satellite systems.

\begin{figure*}
  \begin{center}
	\includegraphics[height=85mm,width=120mm]{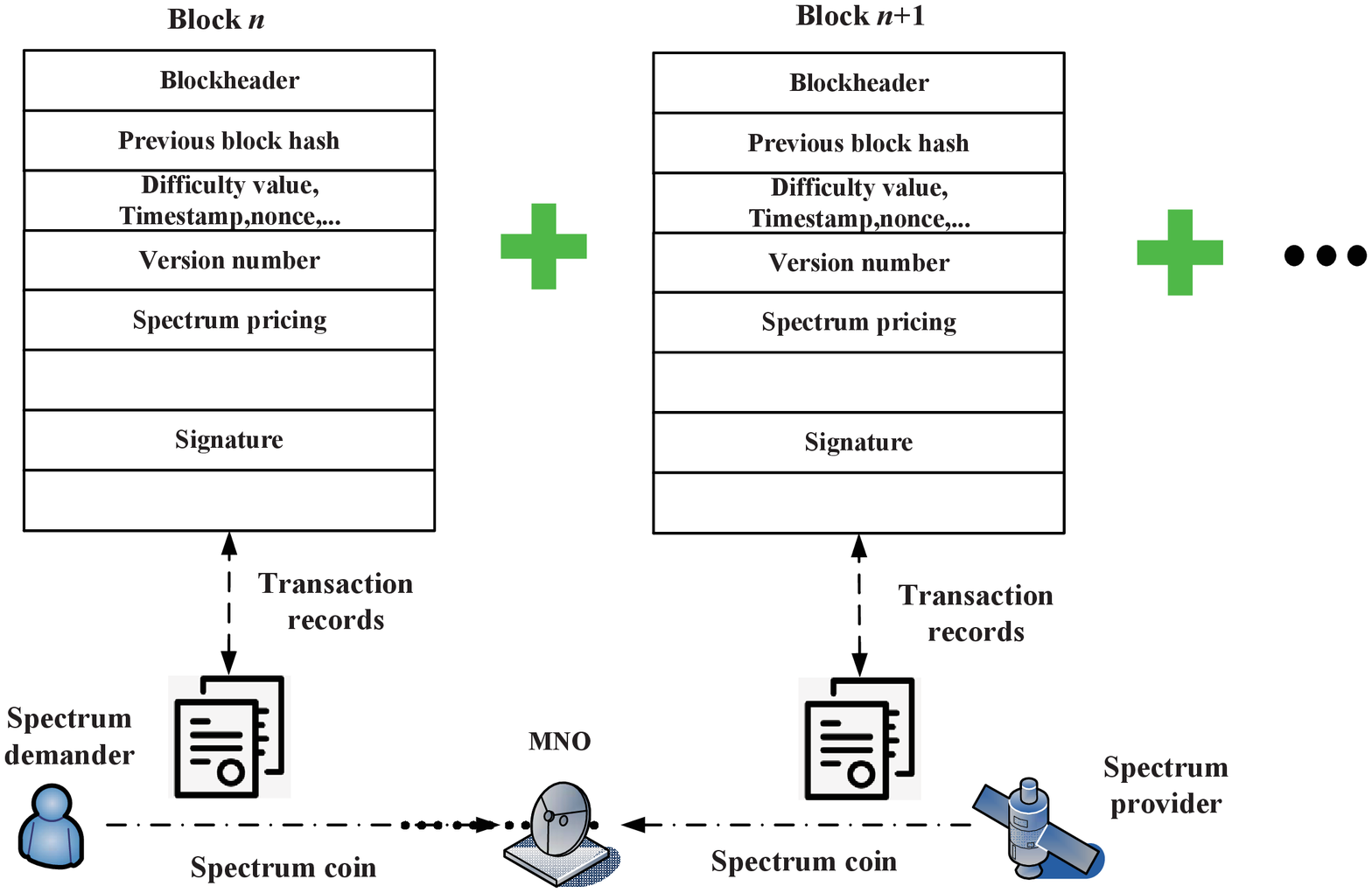}
	\caption{\textbf{Spectrum trading with virtual coin}}
	\label{fig:dummy-figure}
  \end{center}
\end{figure*}

{As shown in Fig. 2, the proposed blockchain-based spectrum trading architecture involves the following components and objects: authorization center, spectrum provider, spectrum demander, virtual coin and terrestrial base station. The authorization center initializes the spectrum trading system, public parameters and cryptographic key. Furthermore, the center sets up initial standard and rule for the blockchain, yet does not participate in specific trading process. In the proposed architecture, edge nodes, which are the terrestrial base stations, are introduced to complete mining calculation, verify the trading users and conduct the spectrum trading. The scheme also introduces an object called spectrum coin which is stored in edge nodes for virtual transaction. In the spectrum trading system, transaction records are saved in each edge node to ensure transparency and communication security. In this case, general terrestrial users serve as the spectrum demanders and satellite system serves as the spectrum provider. Every terrestrial user which aims to use the satellite spectrum needs to pay the virtual coin through the edge nodes. Several significant characteristics of this mechanism are as follows.}

\begin{itemize}
  \item Set the edge calculation nodes to solve the calculation force problem in blockchain creation. Focus on the distributed scenario of terrestrial networks, as shown in Figure 2, select the central node or base station in the local area as the terrestrial agent to communicate with the satellites. The calculation capability and energy reserve of the terrestrial agent are relatively sufficient which enable it to undertake important block creation and mining work.

  \item The mechanism of miner selection based on the trading reputation of edge nodes will be established. In order to prevent malicious and low computation force nodes from accessing the satellite spectrum, a corresponding reputation table is built for each potential access node, recording all the registration, transaction and other historical information. The node's application along with its historical information and transaction records are distributed to all authorized nodes for approving. Constantly update and maintain the reputation table, timely verify the legal block miner identity of the node, so as to ensure the elimination of malicious node access.

  \item  Spectrum coin and reasonable mining coefficient are defined. The transaction between the terrestrial user and the satellite systems is conducted through the virtual spectrum currency. Spectrum coin and its transaction records are saved in given edge node. Meanwhile the edge node that successfully accesses the blockchain through mining also receives the incentive of spectrum coin. Define a reasonable mining difficulty coefficient, eliminate the intervention of low computation force nodes, and maintain a system incentive balance.
\end{itemize}

{The spectrum trading process is as shown in Fig. 3. In this case, it is assumed that all the traders are rational and selfish with the intention of pursuing maximal benefits. The spectrum demander---terrestrial users purchase bandwidth through buying the spectrum coin for payment. In every edge node, there is a ballet address in which terrestrial users store the virtual coin and wait the collection of satellite systems. Both participants' trading behaviors and data, including time stamp, block data, block header, signature, etc., will be recorded in the databases of all the block nodes for tracing and verification at any time. Once a block node has malicious or fraudulent behavior, its reputation value will be reduced until its trading qualification is terminated. }

{After the block is authorized to join in the blockchain, to ensure trading security, all the authorized nodes will obtain a copy of all the sharing data of the whole blockchain. Meanwhile, a distributed consensus mechanism is devised to efficiently attain the blockchain consensus of this spectrum trading. In our scheme, the miner head is designed to firstly broadcast block data, time stamp, specific random number $\varphi$ to all the authorized edge nodes for checking. In order to obtain mutual supervision and verification rights, the edge nodes will detect the block data information and broadcast the verification results with their own watermark. After receiving all the verification feedback, the leader will share all relevant data, and approve the new miner block to join the blockchain to complete the increase of the blockchain. At the same time, if some nodes object to the current block change, the leader will delay the increase of the blockchain and repeatedly send updated data to these nodes in subsequent operations to further reach a consensus. }

The core index to judge whether the potential edge node is qualified or malicious is the node's reputation value. In this case, we define a reputation model for terrestrial edge nodes. Consider a spectrum operator $Op_i$ waiting for joining in the spectrum trading and an edge node $Ed_j$. In the process of spectrum trading, the operator and edge node should often contact with each other. The reputation value of this operator $Op_i$ to its corresponding node $Ed_j$ can be expressed as $w_{i\rightarrow j}$. We can also have $w_{i\rightarrow j}:=\{{Tru}_{i\rightarrow j}, {Unt}_{i\rightarrow j}, {Ind}_{i\rightarrow j}\}$, wherein $Tru_{i\rightarrow j}, Unt_{i\rightarrow j}, Ind_{i\rightarrow j}$ falling in $[0,1]$ denote the trusted, untrusted and indefinite variables, respectively. There is $Tru_{i\rightarrow j}+Unt_{i\rightarrow j}+Ind_{i\rightarrow j}=1$. The logical model of node's reputation can be given as
\begin{equation}
\left\{
  \begin{array}{ll}
    Tru_{i\rightarrow j}=(1-Ind_{i\rightarrow j})\frac{N_p}{N_p+N_n},  \\
    Unt_{i\rightarrow j}=(1-Ind_{i\rightarrow j})\frac{N_n}{N_p+N_n},  \\
    Ind_{i\rightarrow j}=1-suc_{i\rightarrow j},
  \end{array}
\right.
\end{equation}
where $N_p$ denotes the positive interaction number, $N_n$ denotes the negative interaction number, $suc_{i-j}$ denotes the quality of communication. Then, we have the expression of edge node's reputation
\begin{equation}
rep_{j}=\sum_{i=1}^{N}Tru_{i\rightarrow j}+\phi Ind_{i\rightarrow j}
\end{equation}
where $\phi$ is the given probability coefficient and $N$ denotes the terminal number communicating with edge node $j$.

\section{Joint Satellite Spectrum and Interference Pricing Algorithm}

Based on the above blockchain-centric spectrum trading framework for multibeam satellite systems, then we propose a specific spectrum trading algorithm in light of spectrum and interference pricing. This algorithm is performed by edge computing nodes whose main tasks involve the collection of interactive information for both traders, designing rational spectrum pricing strategy and promoting the smooth implementation of spectrum trading. According to our system model, the spectrum reuse and inter-cell interference will be considered. Besides, the significant spectrum heterogeneity caused by capacity diversity in various beams needs also to be addressed. As shown in Fig. 1, the terrestrial user $c$ and $n$ share the same spectrum and interference with each other. Thus, the utility function of multibeam satellite systems can be expressed as
\begin{equation}\begin{split}
U_s&=N_i(\pi_i p_c f_{cs}+\varepsilon B_i-\kappa X_{loss})\int g(\theta)d\theta \\
& s.t.~~~~\gamma_n=\frac{p_nf_{ns}}{p_cf_{cs}+N_0(\alpha_n)B_n}\geq \gamma^{tar}
\end{split}\end{equation}
where $N_i$ denotes the channel number in type $i$, $\pi_i$ denotes the interference pricing, $p_c$ denotes the transmit power of terrestrial user $c$, $f_{cs}$ denotes the channel fading coefficent, $B_i$ denotes the bandwidth, $\varepsilon$ and $\kappa$ denote the monetary coefficient, $X_{loss}$ denotes the marginal cost. $g(\theta)$ denotes the probability density function of the user $c$ with preference parameter $\theta$.

Then, the utility function of terrestrial user requiring to access the satellite spectrum can be given as
\begin{equation}\begin{split}
U_c=&\omega \log_2(1+\frac{p_cf_{cs}}{p_nf_{ns}+N_0(\alpha_c)B_c})-\varepsilon B_i-\pi_i \theta p_c f_{cs}   \\
& s.t.~~~~ \gamma_c=\frac{p_c f_{cs}}{p_n f_{ns}+N_0(\alpha_c)B_c}\geq \gamma^{tar}
\end{split}\end{equation}
where terrestrial user $c$ and $n$ share the same spectrum band as shown in Fig. 1, thus leading to inter-cell interference. $\gamma_c\geq \gamma^{tar}$ means the quality of communication for new user $c$ should be guaranteed. $\omega$ and $\varepsilon$ are monetary coefficients transforming system benefits to uniformed monetary profits in unit of spectrum coin. $N_0(\alpha_c)B_c$ denotes the background noise in this circumstance. The terrestrial user $c$ not only needs to pay for the spectrum cost, but also the interference pricing. Solving the above utility functions, the optimal spectrum pricing and maximal system profits can be obtained.

\section{Numerical Results}

\begin{figure}
  \begin{center}
	\includegraphics[height=65mm,width=90mm]{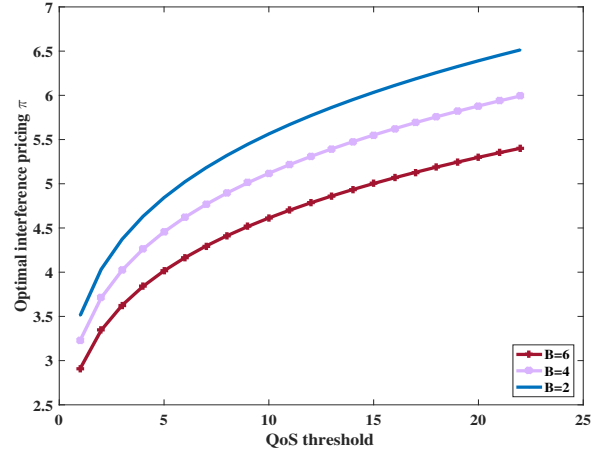}
	\caption{Optimal interference pricing}
	\label{fig:dummy-figure}
  \end{center}
\end{figure}

\begin{figure}
  \begin{center}
	\includegraphics[height=65mm,width=90mm]{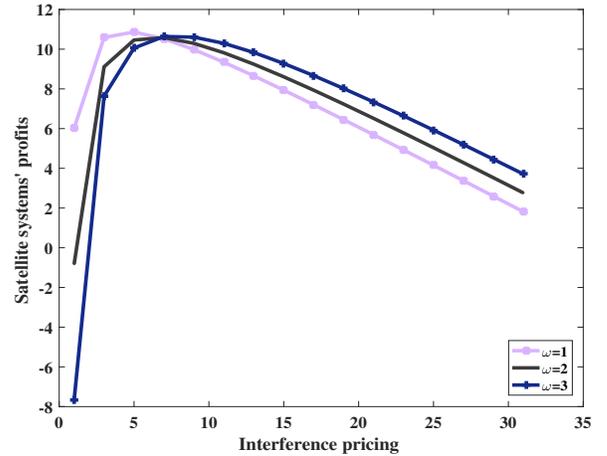}
	\caption{Satellite systems' profits}
	\label{fig:dummy-figure}
  \end{center}
\end{figure}

In this section, numerical results are provided to evaluate the performances of the proposed spectrum trading algorithm in terms of interference pricing and systems' profits. As shown in Fig. 3, we give the trend of optimal interference pricing with changing QoS threshold $\gamma^{tar}$ and leased bandwidth $B$. In this test, we set the parameters as $\omega=1$, $\kappa=0.5$, $N_0(\varepsilon_{Mn})=0.002$, $f_{cs}=0.2$, $f_{ns}=0.1$, $N_i=5$, $N_0(\alpha_c)=0.001$. We can obtain from Fig. 4 that the interference pricing will increase with upgrading QoS threshold $\gamma^{tar}$ which means better communication requirement is needed. As high inter-cell interference will lead to the decline of QoS, the satellite systems require to raise interference pricing to combat with the interference. In addition, we can achieve from Fig. 4 that more available satellite spectrum will decrease the interference pricing. More leased satellite bandwidth can loose the pressure of QoS demand and network capacity.

In Fig. 5, we give the performances of satellite systems' profits with changing interference pricing $\pi$ and monetary coefficient $\omega$. We can achieve from Fig. 5 that excessive interference pricing will reduce satellite systems' profits rather than enhancing profits, since the terrestrial users' communication demands degrade in case of high cost. In addition, as shown in Fig. 5, higher monetary coefficient can benefit satellite systems' profits due to more terrestrial users' benefits in this case.

\section{Conclusions}

In this article, we proposed a blockchain-based spectrum trading algorithm in multibeam satellite systems to enhance spectrum efficiency while guaranteeing satellite communication security. The main contribution of this paper lies in that we designed a satellite spectrum trading method by considering interference pricing and stochastic terrestrial users' spectrum demand while guaranteeing the trading security by using the blockchain technology. To adapt to the specific satellite communication circumstances, we introduce the concepts of edge computing nodes and reputation-oriented node selection mechanism in process of building the blockchain. Besides, due to the spectrum reuse in multibeam satellite systems and heterogeneous transmission in various beams, we take into account the interference pricing and spectrum demand diversity when devising the satellite spectrum trading scheme. Numerical results are also provided to testify the performances of satellite systems' benefits and interference pricing with various key parameter settings.


\end{document}